\documentclass[lettersize,journal]{IEEEtran}
\usepackage{amsmath,amsfonts}
\usepackage{algorithmic}
\usepackage{algorithm}
\usepackage{array}
\usepackage{subcaption}
\usepackage{textcomp}
\usepackage{stfloats}
\usepackage{url}
\usepackage{verbatim}
\usepackage{graphicx}
\usepackage{cite}
\usepackage{colortbl}

\newcommand{\best}{\cellcolor{tablered}}
\newcommand{\sbest}{\cellcolor{orange}}
\newcommand{\tbest}{\cellcolor{yellow}}

\definecolor{yellow}{rgb}{1, 1, 0.7}
\definecolor{orange}{rgb}{1, 0.85, 0.7}
\definecolor{tablered}{rgb}{1, 0.7, 0.7}

\begin{document}

\title{Differentiable High-Performance Ray Tracing-Based Simulation of Radio Propagation with Point Clouds}

\author{Niklas Vaara, Pekka Sangi,  Miguel Bordallo L\'opez, \IEEEmembership{Member, IEEE}, and Janne Heikkil\"a, \IEEEmembership{Senior Member, IEEE}
\thanks{This research was supported by the Research Council of Finland (former Academy of Finland) 6G Flagship Programme (Grant Number: 346208), and the Horizon Europe CONVERGE project (Grant 101094831).}
\thanks{The authors are with the Center for Machine Vision and Signal Analysis, Faculty of Information Technology and Electrical Engineering, University of Oulu, Oulu, Finland.}}

\markboth{Journal of \LaTeX\ Class Files,~Vol.~14, No.~8, June~2025}%
{Shell \MakeLowercase{\textit{et al.}}: A Sample Article Using IEEEtran.cls for IEEE Journals}


\maketitle

\begin{abstract}
Ray tracing is a widely used deterministic method for radio propagation simulations, capable of producing physically accurate multipath components. The accuracy depends on the quality of the environment model and its electromagnetic properties. Recent advances in computer vision and machine learning have made it possible to reconstruct detailed environment models augmented with semantic segmentation labels.

In this letter, we propose a differentiable ray tracing-based radio propagation simulator that operates directly on point clouds. We showcase the efficiency of our method by simulating multi-bounce propagation paths with up to five interactions with specular reflections and diffuse scattering in two indoor scenarios, each completing in less than 90~ms. Lastly, we demonstrate how the differentiability of electromagnetic computations can be combined with segmentation labels to learn the electromagnetic properties of the environment.
\end{abstract}


\section{Introduction}

Radio propagation simulations are an important part of research, optimization, and design of modern wireless communications systems. In the deterministic simulations, physical characteristics of the environment are used to predict radio wave propagation. One approach to such simulation is to use ray tracing (RT), which is based on determining propagation paths and their electromagnetic (EM) properties. Advances in GPU computation related to RT have made this approach increasingly efficient, and there are even applications requiring real-time RT performance. One such example is presented in \cite{andrei2025demonstration}, where a robot equipped with a software-defined radio transmits odometry updates dozens of times per second and each movement of the robot requires recomputation of propagation paths.

The accuracy of the environment model significantly influences the correctness of the simulated paths in RT. Considering geometry, traditional methods have often used hand-crafted triangle mesh models, whereas envisioned radio digital twins will utilize multi-modal sensor data to reconstruct 3D replicas of the environment \cite{alkhateeb2023real}. Typically, sensor-based reconstructions are initially in the form of point clouds, which can be mapped to triangle meshes. Both reconstructed point clouds \cite{virk2015simulating, jarvelainen2016indoor, koivumaki2022point, koivumaki2023ray, vaara2025ray} and reconstructed triangle meshes \cite{pang2021gpu, wahl2022wip, suga2023indoor, kamari2023environment, xia2024path, suga2025rgb} have demonstrated applicability as a basis for RT simulations. However, considering the noisiness of point clouds and computational cost of triangle mesh mapping, it becomes more attractive to directly use point clouds in RT. 

The second aspect in environment modeling is the assignment of EM material properties. Material databases available in the literature are relatively limited, one notable example being ITU-R-P.2040 \cite{itu}, which contains a handful of different material parameters for multiple frequencies. Instead of using such predefined materials, recent advancements in machine learning have shown that differentiable RT, as well as neural representations combined with RT \cite{hoydis2024learning, jiang2025learnable} can be used to learn the EM material properties. Combined with point clouds, one could use semantic point cloud segmentation models such as \cite{schult2023mask3d, kolodiazhnyi2024oneformer3d} to automatically label points based on their material and learn their properties using differentiable RT. Moreover, using point clusters to represent a single material may be beneficial over neural representations when the number of channel measurements is small.

RT algorithms are generally categorized into two types: the image method and ray launching \cite{yun2015ray}. The former computes exact paths for each transmitter-receiver (TX-RX) pair separately, while the latter computes approximate paths simultaneously for all TX-RX pairs by uniformly casting rays from a TX and tracing them. Environment-driven ray launching improves path computation efficiency by discretizing the environment with a given resolution and casting rays only towards these elements \cite{lu2018discrete, koivumaki2022point}. Visibilities between all of the elements and antennas are saved in a matrix, which is expensive to compute, and thus, unsuitable for real-time applications. Moreover, the proposed methods use volumetric rays, which typically increases the number of rays to be cast after each intersection.

In this letter, our main contributions are the following: 1) An efficient ray tracer for computing paths consisting of specular reflections, diffuse scattering, and diffractions. To the best of our knowledge, our method is the first ray tracing-based method for simulating propagation paths directly on point clouds that is capable of computing multi-bounce paths in time, which is suitable for many real-time applications. Moreover, it interfaces with Sionna \cite{hoydis2023sionna}, which enables differentiable electromagnetic computations. As our method can utilize reconstructed point clouds directly, it enables the utilization of fully automated environment model construction from multi-modal sensor data. 2) An improved method for computing intersections with point clouds using surface-aligned 2D disks. Our proposed method computes a weighted average over the intersecting disks, where each weight is defined by a Gaussian on the disk plane combined with distance-based attenuation. 3) Visibility matrix-aided path computation, which enables efficient simulation of diffuse scattering, maintains a constant number of rays throughout the simulation, and minimizes the effects of ray divergence. Our code is available at \url{https://github.com/nvaara/NimbusRT}.

\section{Method}
\begin{figure*}
    \centering
    \includegraphics[width=1.0\linewidth]{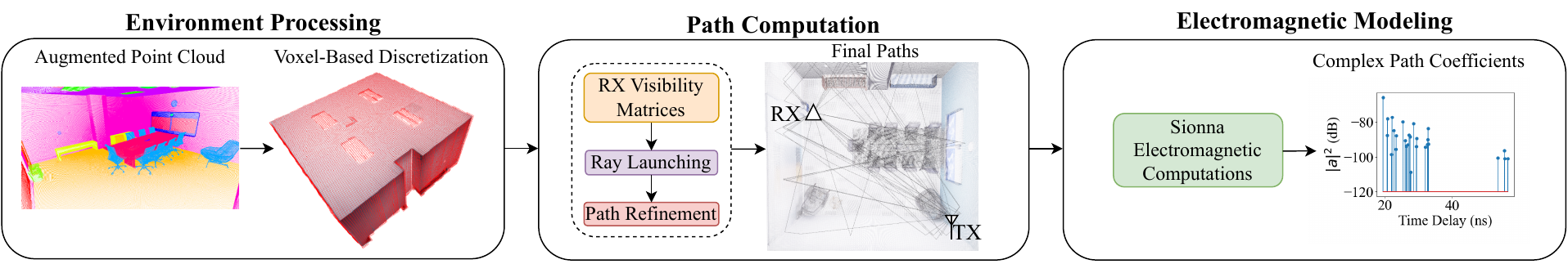}
    \caption{Overview of how the simulator transforms the input point cloud into a ray traceable representation, followed by path and electromagnetic computations.}
    \label{fig:highlevel}
\end{figure*}
Our method takes an augmented point cloud as an input, where each point is associated with a 3D position $\mathbf{p}_i$, surface normal $\mathbf{n}$, surface label $l_{s}$, material label $l_{m}$. In addition, edges can be included for diffraction, described by the edge segment starting point $\mathbf{e}_s$ ending point, $\mathbf{e}_p$, surface normals $[\mathbf{n}_{1}, \mathbf{n}_2]$, and material labels $[l_{e1}, l_{e2}]$ of the adjacent planes and that define the edge. Given a set of TX and RX positions, our method computes paths consisting of specular reflections, diffuse scattering, and diffractions, which can then be passed to Sionna for electromagnetic computations. Our method is generalizable and also supports triangle meshes, however, we focus on point clouds in this letter. An high level illustration of the method is provided in Fig.~\ref{fig:highlevel}.

\subsection{Pre-Processing}

GPU accelerated ray tracing APIs such as OptiX \cite{parker2010optix} utilize acceleration structures for ray traversal. In the case of point clouds, we have to define a set of axis aligned bounding boxes (AABBs) and write our own implementation to determine the intersection point. Similar to our previous work \cite{vaara2025ray}, we discretize the point cloud into voxels with a given resolution $\rho$ and form AABBs for each point set occupying a voxel. Each discretized point set (DPS) also contains a ray reception point, which is the mean position of all the points in the set.

\subsection{Path Computation}

Our path computation begins with visibility matrix construction. Unlike prior work \cite{koivumaki2022point, lu2018discrete, vaara2023refined}, we compute visibility only from each RX to DPS ray reception points, enabling visibility matrix generation in a matter of milliseconds.

The paths are computed for each TX separately. We first cast rays to all RXs to establish the line of sight (LOS) paths. This is followed by casting rays towards all DPSs to determine the initial interaction points with the scene. Diverging from earlier works that utilize environmental information in the ray launching \cite{vaara2025ray, koivumaki2022point, lu2018discrete}, the traced number of rays does not increase. Instead, each interaction is treated as a specular reflection. This allows efficient multi-bounce simulations, with near linear time complexity with respect to the number of bounces. However, the drawback is that the rays are distributed sparsely. In traditional RL this would be problematic, as the rays would need to intersect with a volume around a RX for a path candidate to be found. In our implementation we can mitigate this problem by querying the visibility matrix to determine the path candidates for further processing.

\subsubsection{Diffusely Scattered Paths}

Diffuse scattering can be very expensive to compute as the scattered field spreads across the hemisphere. To ensure fast simulations, we support diffuse scattering as the last interaction point. 
As the $N$ previous interaction points are always specular reflections, the only computation required is the final visibility validation by tracing a ray from the scatter point to the RX. Solely relying on the visibility matrix is not enough, as each DPS in the scene represents a volume rather than a surface.

\subsubsection{Specular Reflection Paths}
\label{spec_ref}
We compute the exact paths similarly to our previous work \cite{vaara2025ray}, which utilizes a path minimization approach derived from the Fermat's principle of least time. However, the refinement can be rather time consuming due to iterative ray tracing operations. To reduce this overhead, we propose a two-stage refinement process.

In the first phase, the length of a path is minimized by optimizing the interaction points on infinite boundless planes as in \cite{vaara2023refined}. Mathematically, the path length can be written as
\begin{equation}
f_{d} = \sum_{k=0}^{N+1} \| \mathbf{I}_k - \mathbf{I}_{k+1} \|,
\end{equation}
where $\mathbf{I}_{k}$ is the $k$th point along the path trajectory. The points $\mathbf{I}_{0}$ and $\mathbf{I}_{N+1}$ are TX and RX, respectively.
We perform optimization iteratively for a maximum of $\xi$ iterations or until convergence. A path is considered to be converged when $ ||\nabla f_{d}||^2 < t_{c}$, where the gradients represent directional derivatives along the plane vectors, which together with the normal vector of the intersection point form an orthonormal basis. In the second phase, each interaction is validated. For the $k$th interaction, this is done by casting a ray from $\mathbf{I}_{k-1}$ to the optimized position. If the angular difference between the previous and new surface normal and the distance to the previous intersection point plane defined by its surface normal are below the thresholds $t_{a}$ and $t_{d}$, the path is considered valid. Otherwise, we retry the optimization $\kappa$ times.

\subsubsection{Diffracted Paths}

Diffracted paths are computationally expensive, as the wave that encounters an edge scatters in multiple directions following the shape of a Keller's cone \cite{keller1962geometrical}. Similarly to diffuse scattering, one could utilize the visibility matrix to validate the visibility from an edge to RXs. However, since Sionna is restricted to single-bounce diffractions, our implementation follows this restriction. As a result, for each edge, we simply minimize the path length similar to specular reflections, as explained in Section \ref{spec_ref}. Lastly, we validate that the edge is visible from both TX and RX.

\subsubsection{Duplicate Path Removal}

As many of the paths converge to the same trajectory, a lot of duplicate paths will be present. We follow the hashing implementation of \cite{vaara2025ray} where the path interaction type and the surface labels $l_{s}$ of the closest points at each intersection are used to form a unique hash per path trajectory. In the case of diffuse scattering we instead use the voxel coordinate for hashing, as each surface may have multiple valid scatter points. Thus, each diffusely scattered path represents the scattered field over an area of size $\rho^2$, approximating the area of the voxel resolution $\rho$.

\subsection{Intersection Computation}

In our previous work \cite{vaara2025ray}, we utilized a signed distance function (SDF) to model the implicit surface of a DPS. 
Finding the zero-level set of the SDF requires iterative ray marching, which may take many iterations to converge. Each point contribution was calculated using the spherically attenuating Gaussian weight function
\begin{equation}
 \delta(\mathbf{s}) = \exp\left(-\frac{||\mathbf{p}_{i} - \mathbf{s}||^2}{2\sigma^2}\right),
\end{equation}
where $\mathbf{s}$ is the sample position at which the SDF is evaluated and $\sigma$ is the attenuation factor. This function does not align well with the actual surface geometry, and the points are not explicitly intersectable. 

To account for these limitations, we propose an alternative approach for computing ray-surface intersections. Each point in a DPS is represented as a planar disk, whose centroid is $\mathbf{p}_{i}$, normal is $\hat{\mathbf{n}}_{i}$, and radius is given by a parameter $r_{p}$. This enables the evaluation of explicit ray-disk intersection $\mathbf{q}_{i}$ for each point in the DPS for a given ray. Finally, the surface intersection point is computed as the weighted average of the intersected disks in the DPS with
\begin{equation}
    \mathbf{q} = \frac{\sum_{i} \mathbf{q}_{i} \omega_{i}}{\sum_i \omega_{i}},
\end{equation}
where $\omega_{i}$ denotes the weight of the $i$th intersected disk. The weights are computed using
\begin{equation}
    \omega_{i} = \exp \left(-\frac{|| \mathbf{q}_{i} - \mathbf{p}_{i}||^2}{2 r_{p}^2}\right) \exp \left( -\lambda_{d} || \mathbf{q}_{i} - \mathbf{q}_{min} || \right),
\end{equation}
where $\mathbf{q}_{min}$ is the ray-disk intersection point closest to the ray origin, and $\lambda_{d} > 0$ is a depth attenuation parameter. The parameter $\lambda_{d}$ can be used to adapt to the noisiness of the point cloud, while $r_{p}$ can be based on the point cloud density.

\section{Experiments}

\begin{figure*}[ht]
    \centering
    \subfloat[The corridor scene and baseline method from \cite{vaara2025ray} are used for performance evaluation. \label{fig:corridor}]{\includegraphics[width=0.4\textwidth]{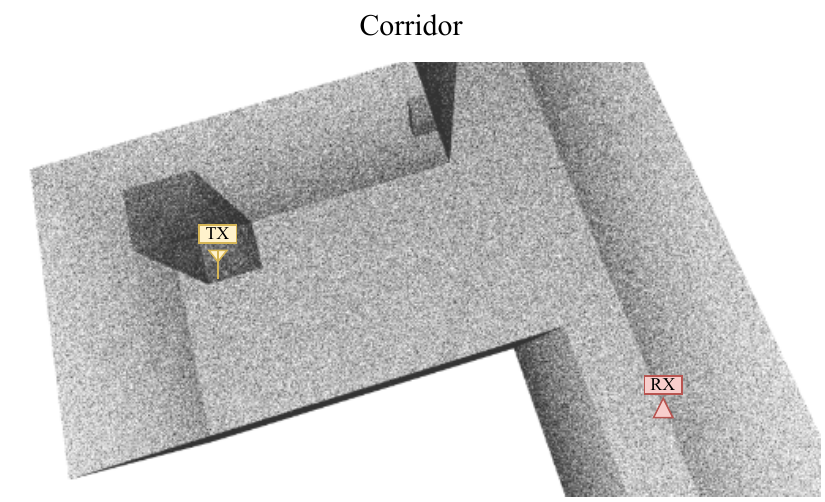} }\hfil
    \subfloat[An indoor scene from Replica dataset \cite{straub2019replica}, which is used for demonstrating the learning of material properties based on material labels. \label{fig:replica}]{\includegraphics[width=0.4\textwidth]{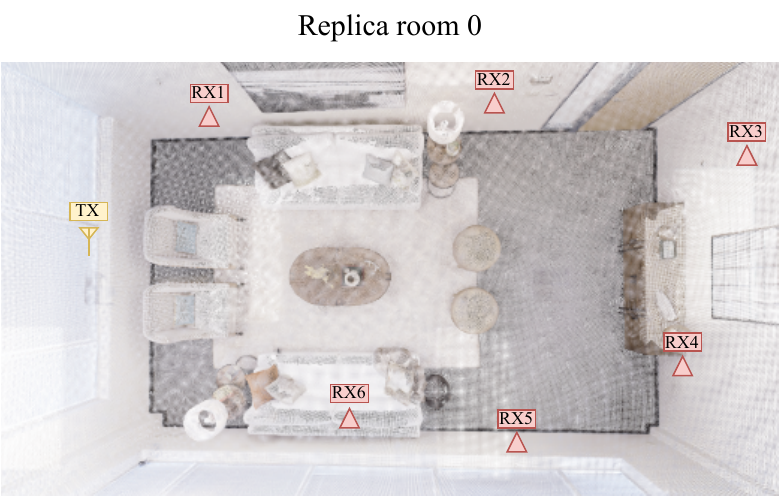} }

\caption{Scenes and antenna positions used in the experiments.}
\label{fig:scenes}

\end{figure*}

In these experiments we demonstrate the efficiency of our simulator and how it can be used to leverage the differentiable EM computations of Sionna. The experiments were conducted on a computer equipped with an NVIDIA GeForce RTX 3080 GPU, and with the parameters presented in Table \ref{tab:params}.

\begin{table}[H]
    \caption{Generic simulation parameters.}
    \centering
    \resizebox{0.7\linewidth}{!}{
    \begin{tabular}{ccc}
         Name & Parameter &  Value\\
         \hline
         Voxel Size & $\rho$ & 0.0625\\
         Point Radius & $r_{p}$ & 0.015 \\
         Distance Attenuation & $\lambda_{d}$ & 100 \\
         Refine Max Iterations & $\xi$ & 50\\
         Refine Retry Iterations & $\kappa$ & 10\\
         Refine Convergence Threshold & $t_{c}$ & 0.0001\\
         Refine Angle Threshold & $t_{a}$ & 1.0 \\
         Refine Distance Threshold & $t_{d}$ & 0.01 \\
    \end{tabular}
    }
    \label{tab:params}
\end{table}

\subsection{Performance Evaluation}

We evaluate our method against \cite{vaara2025ray} and use the same corridor scene (see Fig.~\ref{fig:corridor}) by benchmarking the simulation with 2--5 specular reflections in non-LOS conditions. We use the results achieved with voxel size of 0.5~m from \cite{vaara2025ray} as the baseline, since the smallest discretization size matches our voxel size. From the results shown in Table~\ref{tab:performance} it can be observed that the baseline method \cite{vaara2025ray} starts to slow down rapidly after the number of interactions is increased, while our simulator is capable of finding similar amount of specular paths while maintaining fast execution time. In addition, including diffuse scattering causes only a small computational overhead. 

It should be noted that in \cite{vaara2025ray} refinement was computed with 2000 iterations 
without any convergence-based exit condition and therefore, to have a clearer comparison, the coarse path computation and refinement are separated in Table~\ref{tab:performance}. Our method computes the refined paths after every bounce, while in \cite{vaara2025ray} it is done for all of the coarse paths at once. The implementation of \cite{vaara2025ray} is suited for more complex simulations such as paths consisting of multiple diffractions, while the main focus of the method presented here is rapid simulations. However, it can also simulate single bounce diffractions with negligible overhead.

For further evaluation, we performed a simulation with maximum interaction depth of 5 in the Replica \textit{room 0} shown in Fig.~\ref{fig:replica}, with the illustrated TX position and six RX positions. With specular reflections and diffuse scattering enabled, our method achieved a path computation time of 81--88~ms per link, with an average computation time of 85~ms.

\begin{table*}[b]
\centering
\caption{Performance evaluation of our method with specular reflections and diffuse scattering in the corridor scene.}
\resizebox{0.9\textwidth}{!}{
\begin{tabular}{c|ccc|cc|cc}
 & \multicolumn{3}{c|}{\textbf{Vaara \textit{et al.} \cite{vaara2025ray}, Specular Paths}}
 & \multicolumn{2}{c|}{\textbf{Ours, Specular Paths}}
 & \multicolumn{2}{c}{\textbf{Ours, Specular + Scattered Paths}}\\
\textbf{Max. Depth} & \textbf{Trace (ms)} & \textbf{Refine (ms)} & \textbf{Num. Paths} & \textbf{Trace + Refine (ms)} & \textbf{Num. Paths} & \textbf{Trace + Refine (ms)} & \textbf{Num. Paths} \\
\hline
2 & \tbest 30   & 330  & 9  & \best 10  & 9  & 18 \sbest  & 17~393\\
3 & \tbest 60   & 600  & 25 & \best 15  & 25 & 25 \sbest  & 26~681\\
4 & \tbest 510  & 1100 & 52 & \best 21  & 53 & 37 \sbest  & 36~637\\
5 & \tbest 7650 & 2800 & 90 & \best 28  & 89 & 52 \sbest  & 47~403\\
\end{tabular}
}
\label{tab:performance}
\end{table*}

\begin{figure*}[b]
    \centering
    \includegraphics[width=0.9\linewidth]{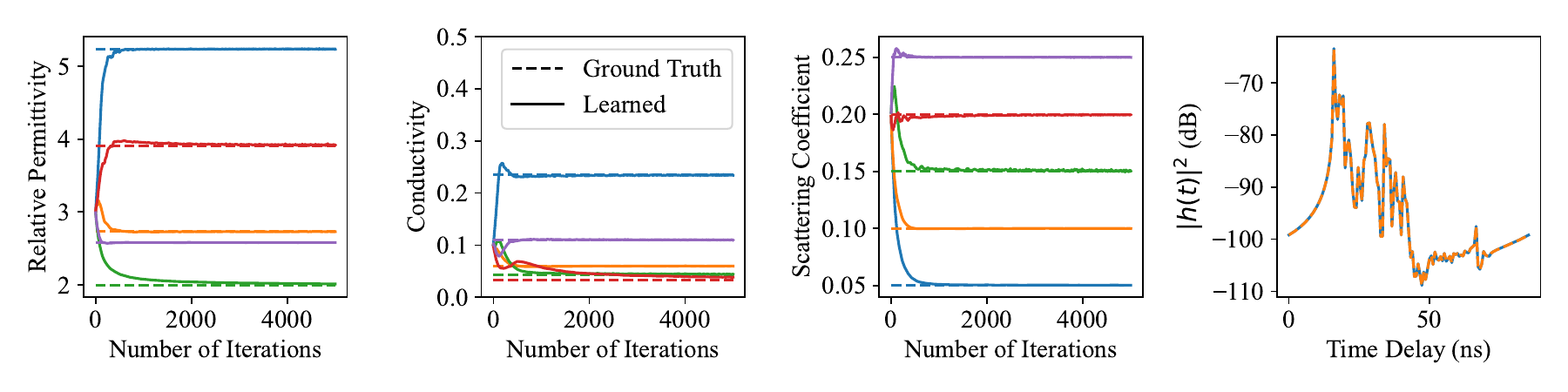}
    \caption{Relative permittivity, conductivity, and scattering coefficient at each training iteration. The rightmost figure shows the ground truth PDP and the PDP computed with the trained materials after 5000 iterations between TX and RX2 in the Replica \textit{room 0} scene (see Fig.~\ref{fig:replica}).}
    \label{fig:learned-mats}
\end{figure*}

\subsection{Learning Electromagnetic Properties from Material Labels}

In this experiment, we showcase how the relative permittivities, conductivities, and scattering coefficients of the material labels could be learned.
For the environment model, the \textit{room 0} scene from Replica \cite{straub2019replica} was chosen. The dataset contains rich geometric information such as class and instance segmentation labels. We use the former labels for the material labels $l_{m}$, while the latter are used as surface labels $l_{s}$. For visualization purposes, we reduce the total number of materials to five.
The paths are simulated up to two interactions, including diffuse scattering with Lambertian scattering model \cite{degli2007measurement}, specular reflections, and LOS at six different RX locations (see Fig.~\ref{fig:replica}). This results in an average of just over 44~000 paths per link. 

As the ground truth for training, channel impulse responses (CIRs) are synthetically generated
by computing channel frequency responses (CFRs) using 129 equally spaced samples over a 1.5~GHz bandwidth centered at 8~GHz. The complex path coefficients are computed at this center frequency. The CIRs and CFRs can be obtained from the path coefficients as shown e.g. in \cite{jarvelainen2016indoor}.

During each train iteration, one of the RX positions shown in Fig.~\ref{fig:replica} is chosen randomly and we compute the CIR from the simulated paths. We use Adam optimizer with a learning rate of $0.01$ and optimize the material parameters with the loss 
\begin{equation}
    \mathcal{L} = \frac{|\mathbf{h} - \mathbf{\hat{h}}|^2}{|\mathbf{\hat{h}}|^2},
\end{equation}
where $\mathbf{h}$ and $\mathbf{\hat{h}}$ are the simulated and ground truth CIR, respectively. Additionally, we apply appropriate activation functions to the optimizable parameters in order to constrain them within their valid ranges. We perform the optimization for 5000 iterations. The results are illustrated in Fig.~\ref{fig:learned-mats}, which showcases the parameters at each training iteration, and that the parameters can be recovered accurately. The rightmost figure illustrates the ground truth power delay profile (PDP) and the respective one with learned parameters. 

\section{Conclusion}

In this letter we presented a ray tracing-based method for simulating radio propagation with point clouds. We demonstrated the efficiency of our method, which was capable of computing multi-bounce paths in less than 90~ms in the experimental cases, having close to linear growth in terms of computational complexity as the interaction depth was increased. We also showcased how the differentiability of electromagnetic computations could be leveraged together with segmentation labels to learn material parameters from ground truth channel impulse responses.

Future work will focus on further experiments with segmented point clouds
constructed from multi-modal sensor data and utilizing real world channel
measurements for material parameter estimation.

\bibliographystyle{IEEEtran}
\bibliography{IEEEabrv,citations}

\end{document}